# How Many Online Workers are there in the World? A Data-Driven Assessment

Otto Kässi[1,2] ✉, Vili Lehdonvirta[2,3], Fabian Stephany[2]

✉ otto.kassi@etla.fi

[1] *ETLA Economic Research*
[2] *Oxford Internet Institute, University of Oxford*
[3] *Alan Turing Institute*

**ABSTRACT**
An unknown number of people around the world are earning income by working through online labour platforms such as Upwork and Amazon Mechanical Turk. We combine data collected from various sources to build a data-driven assessment of the number of such online workers (also known as online freelancers) globally. Our headline estimate is that there are 163 million freelancer profiles registered on online labour platforms globally. Approximately 19 million of them have obtained work through the platform at least once, and 5 million have completed at least 10 projects or earned at least $1000. These numbers suggest a substantial growth from 2015 in registered worker accounts, but much less growth in amount of work completed by workers. Our results indicate that online freelancing represents a non-trivial segment of labour today, but one that is spread thinly across countries and sectors.

## 1 Introduction

Development in digital communication technologies has made transacting work remotely far easier and more economical. At the forefront of this phenomenon are so-called online labour platforms, also known as online outsourcing, crowdwork, or online gig platforms. They allow workers to serve multiple clients at varying hours remotely from their homes or co-working spaces instead of working full-time for a single employer. In this short paper, we refer to the phenomenon as online freelancing, though the employment status of platform-based work is in some cases contested.

Current economic statistics are not well suited to measuring the online freelance economy, in terms of both capturing its full extent as well as distinguishing its impact from other activities (Abraham et al. 2017). Kässi and Lehdonvirta (2018) give several reasons for this: the standard definition of employment is someone who has done at least one hour in the tracking period.

Since online work is often a source of supplementary income (Farrel & Greig, 2018), labour force surveys do not capture it. Moreover, many online workers might not report their earnings to tax agencies, especially if their earnings are small. Tax non-compliance might be particularly prevalent among online workers living in lower-income countries with weaker tax enforcement. In most cases platform companies are not considered employers and thus are not required to report the income earned by the workers (Ogembo & Lehdonvirta, 2020).

Lehdonvirta et al. (2019), Kuek et al. (2015), Horton & Kerr (2015), Braesemann et al. (2018) and Melia (2020), among others, have argued that digital jobs can facilitate virtual migration, or bring jobs to people instead of forcing workers to migrate to where the jobs are. This, in itself, could be a powerful mechanism for development as a large share of the global digital labour force resides in developing countries and social distancing counter-measures against the Covid-19 pandemic bolster the importance of remote online freelancing (Stephany et al. 2020). The argument has been challenged, among others, in Anwar and Graham (2019), Graham and Anwar (2019), Casilli (2017), and Berg, et al. (2019) because digital workers lack formal labour protection and are easily exploited by their employers. We assert that this debate lacks hard data. Since a large share of this activity happens under the radar of national statistical agencies, policymakers and researchers have limited possibilities for assessing the extent and impact of digital labour markets on workers.

The objective of this paper is simple. To assess the global significance of online labour platforms as a source of income, we produce an estimate of the total number of online freelancers globally and document the uncertainties related to the calculation. To facilitate replication and follow-up research, we use publicly available data and make our assumptions explicit.

There are three existing data analyses with a similar goal to ours. Kuek et al. (2015) and Codagnone et al. (2016) used a combination of expert interviews and data disclosed by online labour to estimate the numbers of platform workers globally. Heeks (2017) used estimates from these papers in conjunction with survey estimates to make inferences about the geographic distribution of online work.

Calculations based on expert interviews are useful, but their sources and methods lack transparency and are difficult to repeat regularly in a way that would produce comparable statistics over time. Kässi and Lehdonvirta (2018) took a different approach by estimating the growth rates and geographic distribution of online freelancing by observing vacancies posted on selected English language platforms, but they were not able to count the absolute number of workers filling these tasks.

In addition to international mapping exercises, there have been several national surveys that have assessed the local relevance of digital platform work. These include, among others, Pesole, et al (2018) and Huws et al (2017) who both concentrate on selected European countries. Unfortunately, many populous countries that supply large shares of online labour have not completed such surveys. Moreover, many surveys fail to distinguish local platform work, and activities such as e-commerce and house rental from remote platform work. Another, related, but distinct approach is used in Le Ludec, et al. (2020), and Difallah, et al (2018). These papers used a capture-recapture model inspired by ecological sciences to infer the size of microworker population in France, and number of workers on a single platform, respectively.

The work presented in this paper is thus to our knowledge the first to use a fully quantitative and transparent approach to estimating the absolute number of online workers globally. Beyond providing a headline number, the more general contribution of this paper is that we outline the relevant quantities a researcher needs to know when trying to understand how many online workers there are on the globe.

## 2 Data and methods

We started by attempting to create, as far as possible, a complete census of all online labour platforms of non-trivial size. We used three main sources of information to compile a list of platform names. First, we analysed a publicly available database of the crowd-sourced company information platform Crunchbase, especially its 'freelance' and 'crowdsourcing' categories. Our second data source for platform names is a cross-regional survey collected in Wood et al. (2019). Finally, we supplemented our list with information found through Google searches concerning Spanish, Latin American, Russian, and Chinese online freelancing platforms.

We limited our attention to platforms where the transaction is fully digital; that is, the work is delivered and paid remotely over the internet. Local gig economy platforms such as ride-hailing apps and food delivery platforms are thus excluded. Distinguishing the online freelance economy from the local gig economy is important because, among other reasons, the transnational nature of the market means that it has different potential implications to global service trade and development. We ended up with a list of 351 online freelancing platforms. To best of our understanding, these 351 platforms constitute nearly the full universe of online freelancing platforms in the end of 2020.

We then used public data sources to obtain three measures of worker numbers for each platform: number or registered worker profiles, number of registered workers who have ever worked, and number of registered users who have worked full-time, i.e., who completed at least 10 projects, or earned at least $1000, when available. We collected these numbers through a combination of media mentions, literature review, and platforms' search functionalities.

We were able to observe the number of registered workers in 162 of 351 cases. The distribution of this variable is plotted in Figure 1. The sum of registered workers across all the 162 platforms is 140,000,000[1]. Most of the platforms have fewer than a million registered workers. Three outliers have particularly large numbers of registered workers: freelancer.com (31 million workers), epwk.com (23 million registered workers), and zbj.com (23 million registered workers).

We were able to observe the number of workers who had ever worked on the platform for only 7 platforms. For 6 platforms we were able to observe the number of workers who had worked full time, i.e., who completed at least 10 projects, or earned at least $1000. Fortunately, this information is available for some of the largest platforms, such as Freelancer, as well as for some smaller ones.

---

[1] We round all numbers to two significant digits throughout this paper.

For those platforms for which we could not obtain numbers from public sources, we instead imputed the quantities as described in the next subsection.

## 2.1 Predicting number of registered workers

Previous research has used various rules-of-thumb methods for estimating numbers of workers registered on online labour platforms. For example, Kuek et al. (2015) assumed that the top three firms form 50 % of the entire online freelancing market. Using this assumption, by obtaining data on the top three platforms only, they generalised their findings to the market as a whole.

We instead adopted a data-driven approach to predict the number of workers for the platforms for which this information was not available. We collected a list of publicly available predictive features and trained a machine learning model to predict the number of workers for the platforms where this information was not available.

We use the following predictive features, all measuring different aspects of website popularity, in building the model:

- **Alexa rank.** Alexa is a web traffic analysis company whose data is frequently used to compare the popularity of different websites. We used the most recent Alexa rank as reported by the siterankdata.com analytics tool (accessed 2020-09-29). If Alexa rank is not reported for a give site, we have inputted the maximum in the data as the rank.
- **Estimate for monthly unique users.** Estimated number of monthly unique users as reported by the analytics tool siterankdata.com (accessed 2020-09-29). If the estimate is not available, we have inputted a zero.
- **Median of daily Google Trends index values between 2019-09-01 and 2020-09-01.** Downloaded from Google Trends using the site url (e.g. 'upwork.com') as the search term. If values are not available, we have inputted a zero.[2]
- **Sum of daily Google Trends index values between 2019-09-01 and 2020-09-01.** Downloaded from Google Trends using the site url (e.g. 'upwork.com') as the search term. If values are not available, we have inputted a zero.

The data are summarised in Table 1. Google Trends produces time series data, which for the purposes of this exercise we needed to summarise as a single number. Since there is no one theoretically correct way of doing this, we used two different summarisations: the sum and median of the time series' values, and relied on a machine learning algorithm to choose relevant weights for both features. All features were standardised by subtracting the mean and dividing by the standard deviation and logarithmised before entering them into the model.

Our training data consists of 159 observations. We decided to use 128 observations (80%) as our validation set and 31 (20%) as our training set. After some experimentation, we chose an XGBoost (Chen & Guestrin 2015) model with a Poisson objective. The main upside of the Poisson assumption is that it naturally limits the dependent variable to positive values without any additional adjustments. After an extensive grid search along the model's hyperparameters, we chose the model that minimised the Root Mean Square Error (RMSE) of the prediction in the validation data. The performance of the best performing model is reported in Table 2. We see that the RMSE of the predictions is relatively large at 577,000, which reflects the fact there is non-negligible uncertainty in the prediction. Nonetheless, as we argue below, despite the large RMSE, the prediction is informative of the number of workers registered on platforms.

---

[2] Google Trends is a product that summarises search volume by time. For more details, see: https://trends.google.com/trends (accessed 2020-03-27). A feature of the Google Trends tool is that it is only possible to compare five keywords for each search. To work around this issue, we always pull data for 'upwork.com' and four other platform names, and continuously normalise the search terms as $GoogleTrend_s/GoogleTrend_{Upwork}$. This way, all index values are relative to Upwork.

Adding up the predicted numbers of registered workers across the platforms yields a total of 23,000,000 workers. Adding to this the number of directly observed workers (140,000,000) yields 163,000,000 workers, which is our point estimate for the number of registered workers across the global online freelance economy. Figure 2 plots the distribution of observed and predicted numbers of workers across platforms. We see that the platforms for which we predict the numbers of registered workers are predominantly on the smaller end.

To indicate the uncertainty related to the prediction, we have also estimated a 95% prediction interval for the numbers of registered workers by bootstrapping.

## 2.2 Inferring the number workers who have worked

On most online labour platforms, the number of registered users might not represent the number of online workers who are actually active. To capture this, we follow a similar approach as above and generalise from the known population.

Figure 2 plots the ratio of workers who have ever worked to the total number of worker profiles on those platforms for which this information is known. Figure 3 analogously plots the ratio of workers who have earned at least $1000 or completed at least 10 projects in their careers to the total number of worker profiles. Since only a handful of platforms reveal this information publicly, the sample sizes for these estimates remain very small. Thus, instead of calculating a formal confidence interval for these estimates, we use the minimum and maximum values of the samples as our error band estimate in sensitivity analyses.

The main takeaway from Figures 2 and 3 is that the share of registered workers who do any work is relatively small on the platforms we observe. On average, only 5 % of registered users have ever worked, and 4 % have worked over the past 30 days preceding the data collection. These findings support the notion that for a large majority of the platform workforce the platform is a source of occasional additional income rather than the main income source.

## 2.3 Multi-homing

Multi-homing, or the practice of agents being affiliated with more than one platform, can lead to double-counting of workers. If a worker is active or registered on more than one platform, they will be counted more than once in our data.

There are no measures for double-counting available through public data sources. Fortunately, questions about multi-homing have been asked in several surveys administered by the ILO (ILO 2021), and by Wood et al. (2019a). These surveys asked active freelancers to list how many platforms they worked on. Across the surveys, 48% of the respondents mentioned that they worked exclusively on a single platform. On average, the survey respondents were active on 1.83 platforms. Thus, we can further adjust down the number of active worker profiles to account for multi-homing by dividing the number of active workers by 1.83.

However, qur results on multi-homing could be challenged because they our results on multi-homing are from a non-representative convenience sample. Nonetheless, we note that our numbers align well with those reported in Le Ludec, et al. (2020).

### 2.3 Multi-working: multiple workers using a single account

Qualitative evidence discussed in detail in Lehdonvirta et al. (2015), Wood et al. (2019b) and Melia (2020) suggests that in some cases several workers might be working under a single freelancer account (multi-working). To the best of our knowledge, there are no systematic studies on this phenomenon.

The surveys discussed in ILO reports and in Wood et al. (2019b) asked the following three questions from workers:
"*Over the last 7 days, I have hired workers in my local area to do online work that I got from a client*", "*Over the last 7 days. I have hired family or friends to do online work that I got from a client*", or "*Have you ever participated in digital platform work using a login / account / profile that belongs to someone else or that is shared by multiple people?*"

21% of respondents across the surveys answered yes to one of these questions. If we further assume that an account is shared between a maximum of two workers, we can adjust our numbers for multiple workers working under a single account by multiplying the number of workers by 1.21.

# 3 Results

This section combines the individual parameter estimates discussed above into a single number. Moreover, we provide data-driven error bands for the parameters underlying our estimate.

According to these numbers, our point estimates suggest that there are 163 million registered worker profiles online freelancing platforms. Of them, roughly 19 million have ever worked, and 5 million have had a full-time job (full time job defined as total earnings of over $1000, or over 10 completed projects). Further adjusting for multi-homing, these numbers reduce to 10 million and 2.7 million respectively. Finally, adjusting for possible multi-working increases these numbers to 8.5 million and 2.3 million respectively.

Nonetheless, there is considerable uncertainty in these estimates. Given the relatively large error bands, our estimates suggest that there could at most be as many as 205 million registered worker profiles, 24 million workers who have ever worked through an online labour platform, and 6 million full time workers.

# 4 Discussion and conclusions

This paper used a combination of data sources to produce an estimate for the number of online workers in the online freelancing economy.

According to our headline estimates, we estimate that there are 163 million registered workers on online labour platforms, and 11% of them have ever worked through a platform. These numbers point to a stark growth if compared with the 2015 estimates by Kuek and colleagues (2015), whose corresponding numbers were 50 million and 10%.

The differences in these estimates are not only due to methodological differences. Kuek and colleagues assumed that Upwork, Freelancer, and Zbj form half of the total market. Using their assumptions with our data, our estimate would have been 130 million freelancers in 2020.

Instead, new platforms with a large reach have emerged between 2014 and 2020. Moreover, it could be the case that there are now more, and more geographically or professionally specialised online labour platforms than when Kuek and colleagues (2015) conducted their study.

At face value, our estimates support the narrative that online work is growing rapidly (Huws, et al. 2017, Chan & Wang 2018). Nonetheless, the fact that only a small minority have completed any projects, let alone a substantial number of projects suggest that digital platform work is a viable way to make a living only to a small minority of registered workers.

Nonetheless, we stress that our estimates come with fairly big error bands. Nonetheless, even the lower end of our estimates suggest that the online freelancing economy has grown. For instance, the Online Labour Index (Kässi & Lehdonvirta, 2018) indicates yearly growth rates of over 10%. Upwork, one of the larger online labour platforms, has reported almost 20% year-on-year growth rates in gross freelancer revenues[3].

We believe that our approach is transparent and our methodological choices sound. Nonetheless, there are a few sources of error that could bias our estimates downwards that we cannot tackle. In particular, we want to highlight two error sources.

First, the estimates for the shares of active workers, multi-homing, and multi-working come from small opportunistic samples from limited countries. It is very possible that multi-homing and account sharing practices vary considerably by country and platform.

Second, quantitative evidence on the extent and nature of working on shared accounts (multi-working) is particularly slim. For more reliable evidence on these, we would need representative surveys of freelancers working on the major platforms. Fortunately, since only a handful of platforms cover most of the market, a survey that covers only the major platforms should give us a good understanding of the total market.

More broadly, platform mediated remote work is just one facet of computer-mediated labour. Other facets, such as platform-mediated place-based work, i.e., the local gig economy of ride hailing and delivery services, remote work for overseas clients, and business process outsourcing, can have their own specific impacts on economic development, work and labour

---

[3] https://investors.upwork.com/news-releases/news-release-details/upwork-reports-second-quarter-2019-financial-results/ (accessed 2019-11-08)

market statistics. We hope to see more research on developing better measures for these phenomena as well.

# Figures and Tables

**Figure 1 - Number of freelancers:** On 45 out of 151 registered platforms, we are able to observe the number of registered freelancers.

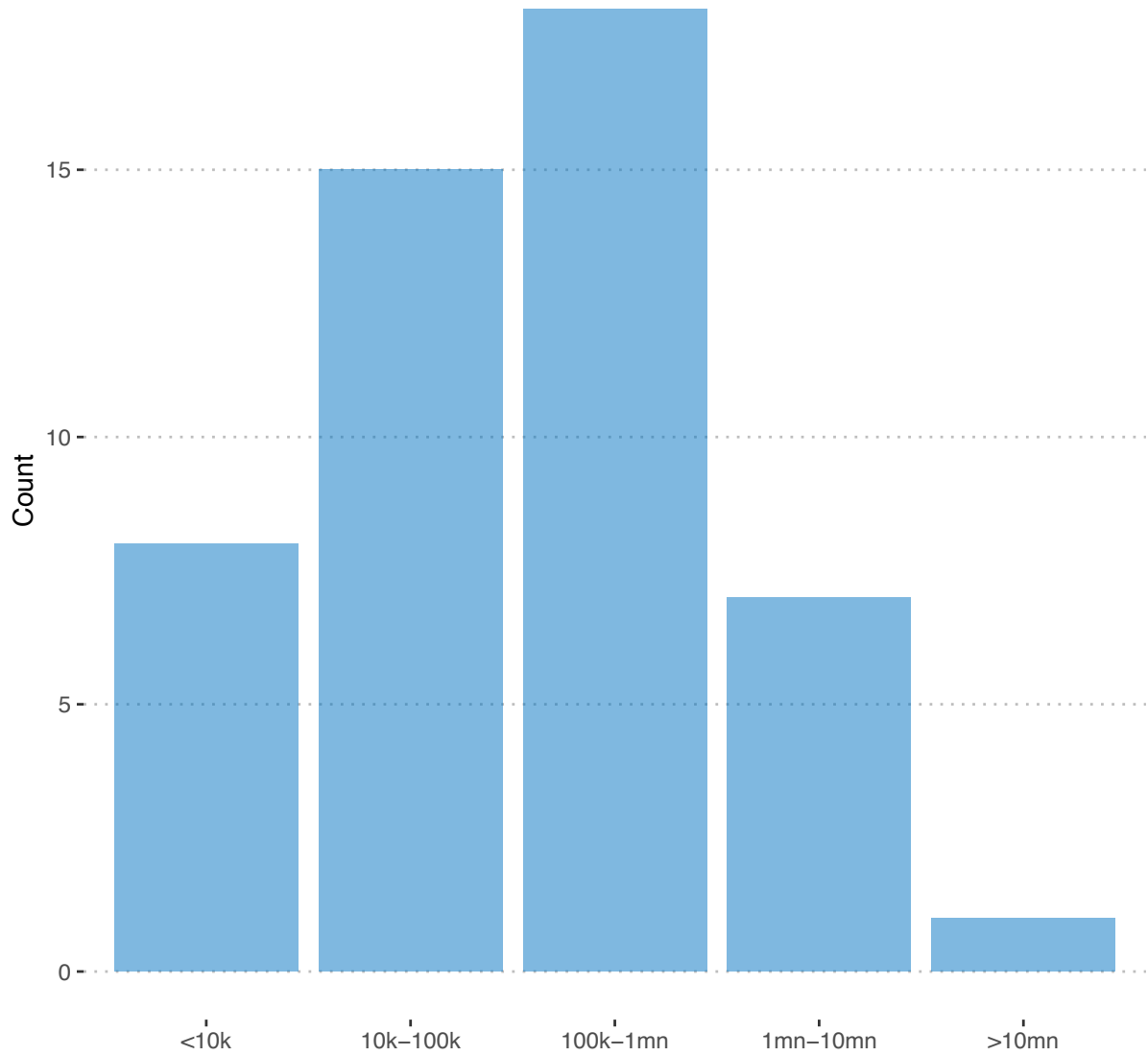

**Figure 2 - Number of freelancers:** Distribution of registered workers for predicted and observed platforms.

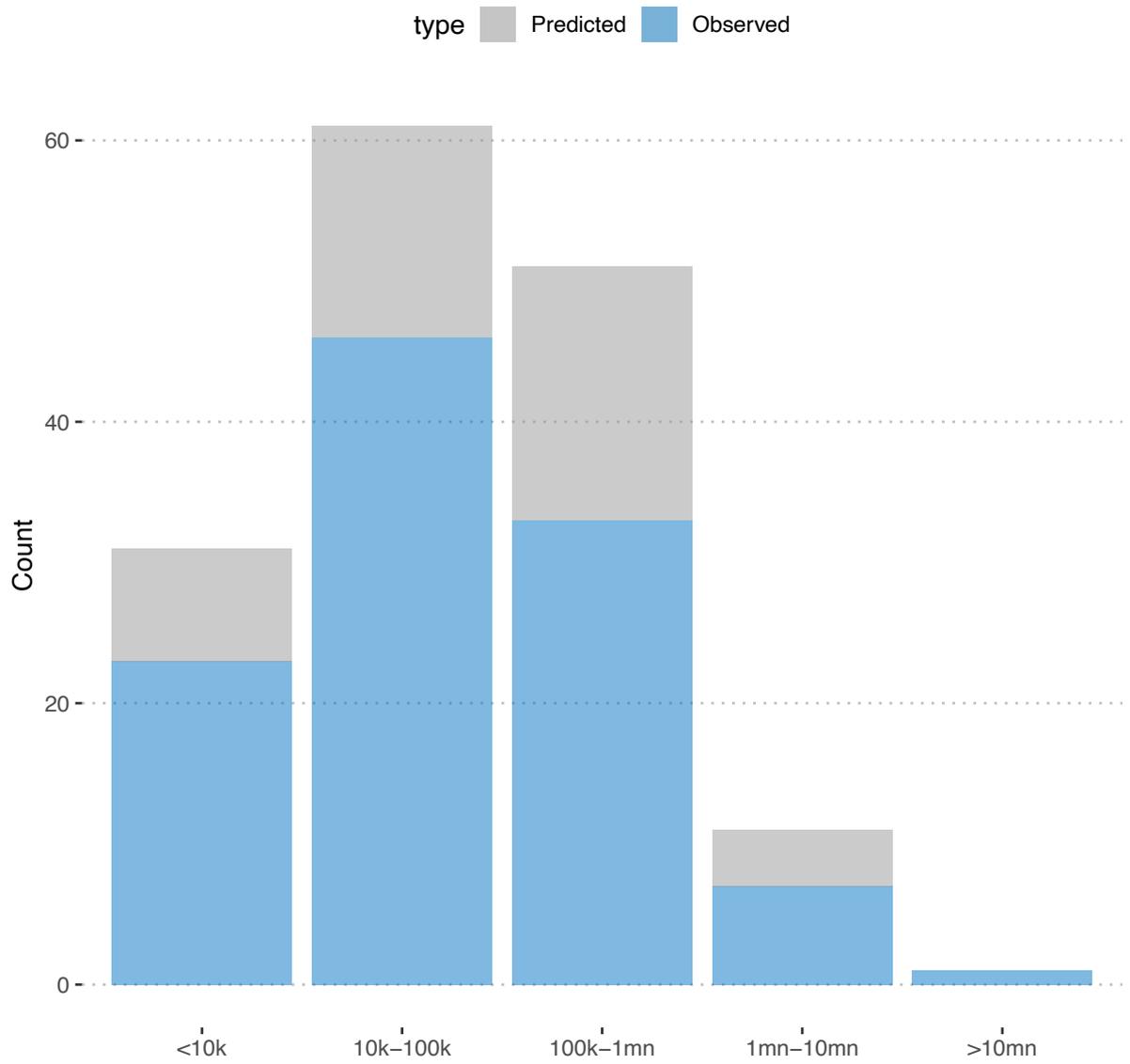

**Figure 3 - Number of active freelancers.** Notes: the graph shows the share of reported registered workers who have completed at least one project. Red line corresponds to average in data.

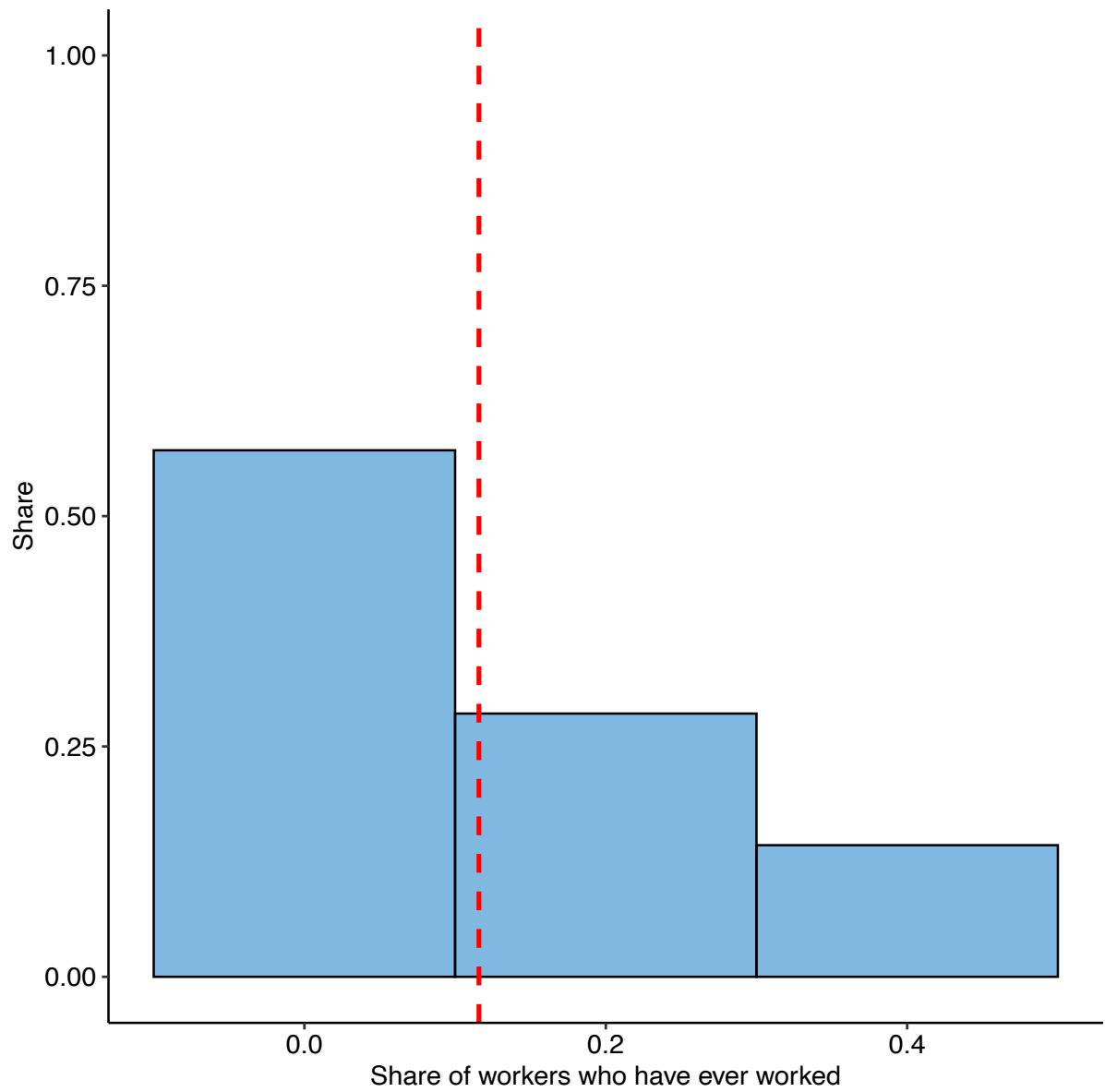

**Figure 4 - Number of recently active freelancers.** Notes: the graph shows the share of workers who have either completed 10 projects, or earned $1000. Red line corresponds to mean in data.

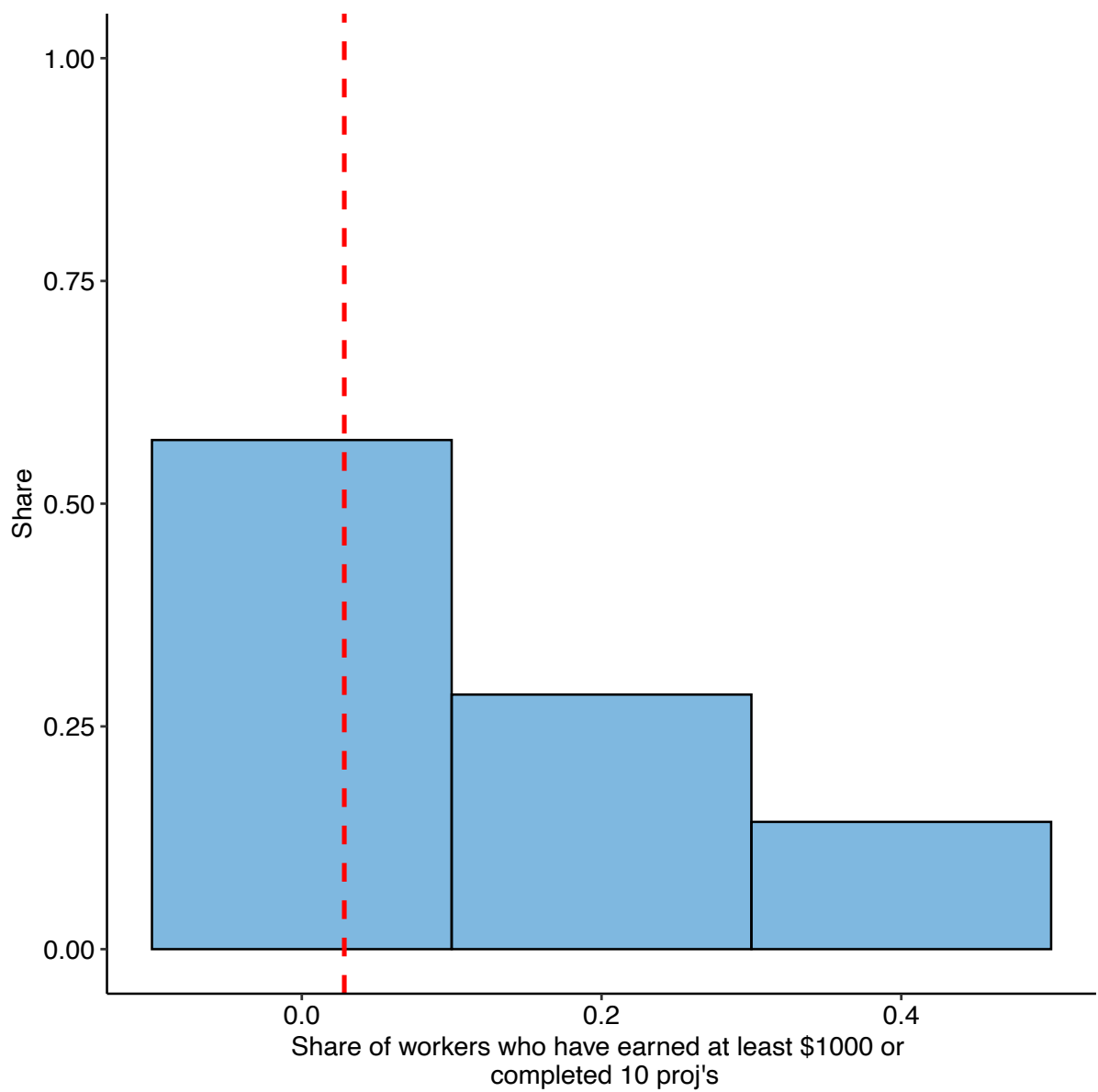

## Table 1 - Summary statistics (observed data)

|  |  | N | Mean | Median | Min | Max |
|---|---|---:|---:|---:|---:|---:|
| **siterankdata.com** | | | | | | |
| | Alexa Rank | 339 | 282,182 | 258,266 | 346 | 986,677 |
| | Estimated monthly unique users | 339 | 4,697,358 | 367,782 | 95,241 | 289,444,543 |
| **Google trends** | | | | | | |
| | Sum of index values | 351 | 0.02 | 0.01 | 0.00 | 1.40 |
| | Median of index values | 351 | 0.02 | 0.00 | 0.00 | 1.59 |
| **Number of Workers** | | | | | | |
| | Total | 162 | 869,173 | 50,000 | 80 | 31,464,473 |
| | <10k | 37 | 3,237 | 2,700 | 80 | 8,000 |
| | 10k-100k | 57 | 33,648 | 25,187 | 10,000 | 90,237 |
| | 100k-1mn | 45 | 321,628 | 259,508 | 100,000 | 856,327 |
| | 1mn-10mn | 19 | 1,891,763 | 1,480,933 | 1,000,000 | 7,000,000 |
| | >10mn | 4 | 22,087,877 | 22,443,518 | 12,000,000 | 31,464,473 |
| | Ever worked | 7 | 13,959 | 8,643 | 2,217 | 36,754 |
| | Worked last month | 6 | 3,724 | 2,828 | 345 | 10,798 |

**Table 2 - Estimation results.** While observing 140 million workers, we predict there to be an amount of 23 million unobserved freelancers worldwide. Notes: For row (b), the error band is calculated as bootstrapped 95% confidence interval; for rows (c), and (d) the error bands are calculated as minimum and maximum values in the data; for row (e) the error band is calculated as the 2.5[th] and 97.5[th] percentile point of the distribution of platforms mentioned; for row (f) the error band is calculated as +/- 1.96 * std. error. All numbers are rounded to two significant digits. See text for details.

|     |                                              | Estimate    | Error band                  |
|-----|----------------------------------------------|-------------|-----------------------------|
|     | Number of workers                            |             |                             |
| (a) | registered (observed)                        | 140,000,000 |                             |
| (b) | registered (predicted)                       | 23,000,000  | [12,000,000, 65,000,000]    |
| (c) | with at least one project completed          | 11.6%       | [0.5%, 36.8%]               |
| (d) | with 10 completed projects or $1000 earned   | 3%          | [0.1%, 10.0%]               |
| (e) | Average of multi-homing platforms            | 1.83        | [1, 4]                      |
| (f) | Proportion of workers sharing accounts       | 0.21        | [0.18, 0.24]                |